\documentclass[aps,prd,preprint,endfloats,showpacs,superscriptaddress,nofootinbib]{revtex4}
\usepackage{amsmath,amssymb,amsbsy,color}
\usepackage{moreverb}
\usepackage[dvips]{graphicx}
\usepackage{psfrag}
\usepackage{indentfirst}
 
\begin{document}
 
\draft
\preprint{UTHEP-556}
\preprint{UTCCS-P-39}
\preprint{KEK-CP-209}
 
\title{Full QCD calculation of neutron electric dipole moment with the external electric 
field method}
\author{E.~Shintani}
\affiliation{High Energy Accelerator Research Organization (KEK), Tsukuba 305-0801, Japan}

\author{S.~Aoki}
\affiliation{
Graduate School of Pure and Applied Sciences, University of Tsukuba,
Tsukuba, Ibaraki 305-8571, Japan }
\affiliation{
Riken BNL Research Center, Brookhaven National Laboratory, Upton, 11973,
USA}

\author{Y.~Kuramashi}
\affiliation{
Graduate School of Pure and Applied Sciences, University of Tsukuba,
Tsukuba, Ibaraki 305-8571, Japan }
\affiliation{
Center for Computational Sciences, University of Tsukuba, Tsukuba,
Ibaraki 305-8577, Japan }

\date{\today}

\begin{abstract}
We have calculated the neutron electric dipole moment (NEDM) in the
presence of the CP violating $\theta$ term in lattice QCD with 2-flavor 
dynamical clover quarks, using  the external electric field method. 
Accumulating a large number of statistics by the averages over 16
different source points and over forward and backward nucleon
propagators, 
we have obtained non-zero signals of neutron and proton EDM
beyond one standard deviation at each quark mass in full QCD.
We have investigated the quark mass dependence of nucleon EDM in full QCD,
and have found that
nucleon EDM in full QCD does not decrease toward the chiral limit, 
as opposed to the theoretical expectation.
We briefly discuss possible reasons for this behavior.
\end{abstract}
\maketitle

\section{Introduction}
A requirement for renormalizability in QCD allows a CP violating term with a free parameter $\theta_{\rm QCD}$,  thus called the $\theta$ term,
\begin{equation}
  \mathcal L_{\delta\rm CP} = \theta_{\rm QCD}\frac{g^2}{32\pi^2}G_{\mu\nu}\widetilde G^{\mu\nu},
  \label{eq:lag_cp}
\end{equation}
where $G_{\mu\nu}$ and $\widetilde G_{\mu\nu}$ represent a gluon field strength and its dual, and 
$g$ is a coupling constant of QCD.
If the CP invariance  is preserved in the strong interaction in Nature,
$\theta_{\rm QCD}$ must be zero in QCD.
Indeed a current experimental bounds of the neutron electric dipole moment (NEDM)
\cite{Harris},
\begin{equation}\label{eq:dn_exp}
   d_N < 6.3\times 10^{-13} {\rm e \cdot fm},
\end{equation}
and the crude theoretical estimates\cite{model_CA,CHPT_dn,QCD_sum},  $d_N = \mathcal O(10^{-2})\theta_{\rm QCD}$,
gives the constraint  that $\theta_{\rm QCD}\lesssim\mathcal O(10^{-10})$.
$\theta_{\rm QCD}$ must be very small or even zero.

In the presence of the electroweak interaction, however, the above conclusion on $\theta_{\rm QCD}$
is modified.
In the Weinberg-Salam(WS) model of the electroweak interaction,  
the quark mass term generated through Yukawa couplings by the spontaneous 
electroweak symmetry breaking is given by 
\begin{equation}
  \mathcal L_{\rm m} = \bar q^i_L(M_{\rm CKM})_{ij}q_R^j + {\rm h.c.},
\end{equation}
with a quark mass matrix $M_{\rm CKM}$ and left- and right-handed quark fields $q_{R,L}$.
In order to transform the quark mass matrix to a real and diagonal form,
U(1) as well as SU($N_f$) chiral rotations are necessary, 
since $M_{\rm CKM}$ is non-Hermitian in general.
Through Adler-Bell-Jackiw anomaly, 
this U(1) chiral transformation shifts $\theta_{\rm QCD}$ in eq.(\ref{eq:lag_cp}) to
\begin{equation}
  \bar\theta = \theta_{\rm QCD} + \arg\det M_{\rm CKM}.   
\end{equation}
Therefore the experimental bound (\ref{eq:dn_exp}) on $d_N$ leads to $\bar\theta \lesssim\mathcal O(10^{-10})$:
A subtle cancellation between a parameter in QCD ($\theta_{\rm QCD}$) and a phase of mass matrix 
in the WS model ($\arg\det M_{\rm CKM}$) should be fulfilled to keep the CP invariance in the strong interaction. 
This cancellation seems unnatural and thus a new mechanism must exist to explain 
the smallness of $\bar\theta$ ("strong CP problem").

A simplest solution to the strong CP problem is that one of quarks is massless, 
so that we can set $\bar\theta=0$ by the chiral rotation of this quark without introducing complex 
phases to other quark masses.
Detailed analyses by CHPT \cite{GL} or lattice QCD \cite{cppacs_jlqcd}, however, 
strongly indicate that up quark has a finite mass ($1.5<m_u<3.0$ MeV \cite{PDG}), 
therefore this solution is excluded. 
In the Peccei-Quinn (PQ) mechanism \cite{PQ},
$\bar\theta$ is promoted to a scalar field associated with a new symmetry, called PQ symmetry,
which is slightly broken by the anomaly.
The spontaneous PQ symmetry breaking automatically pick up an unique vacuum, in which $\bar\theta$ vanishes.
As a consequence of the symmetry breaking,
a very light new scalar particle, called axion must exist in this mechanism. 
So far the axion, which is also one of the candidates for the cold dark matter,
 has not been observed yet, and  
both cosmological observations and accelerator experiments set a very narrow allowed region
of the axion mass such that $10^{-6} < m_a < 3\times 10^{-3}$ eV \cite{axion_rev}. 

An aim of our investigation in this paper is NOT to propose a new solution to the strong CP problem, 
but is to establish a reliable way of calculating the NEDM in lattice QCD. 
For small $\bar\theta$, the NEDM is proportional to $\bar\theta$ as 
$d_N = d_N^{(1)}\bar\theta + O(\bar\theta^3)$. 
Various model calculations \cite{CHPT_dn,model_CA} lead to different estimates, ranging that
$\vert d_N^{(1)}\vert  = \mathcal O(10^{-2}\sim 10^{-3})$ e$\cdot$fm, and 
even its sign has not been determined yet.
The lattice QCD has a potential to calculate $d_N^{(1)}$ non-perturbatively and in a model independent way.
Once a method of calculating the NEDM is established in lattice QCD for the case of a particular 
CP violating term given in (\ref{eq:lag_cp}),   
it may become possible to extend NEDM calculations to the case of new CP violating terms such as
the chromoelectric dipole moment \cite{QCD_sum} generated in SUSY models at high energy\cite{Pospelov:05}. 

Lattice calculations of $d_N^{(1)}$ have remained to be notoriously difficult for a long time \cite{Aoki}.
In our previous papers \cite{Shintani_form, Shintani_wE} we have investigated two methods of
calculating $d_N^{(1)}$ in quenched lattice QCD.
In the first method \cite{Shintani_form}, 
we have calculated the NEDM form factor at non-zero momentum transfer, which becomes the NEDM in the limit of zero momentum transfer.
In the second method, the NEDM has been directly extracted from the energy shift in the presence of 
the external electric field. In both cases, we have successfully obtained non-zero signals of EDM 
for neutron and proton.
In this paper we extend the calculation of the NEDM to the case of 2-flavor full QCD, using configurations
generated by the CP-PACS collaboration \cite{cppacs}.
We employ the second method, the direct calculation of the NEDM with the external electric field,
since no extrapolation of momentum transfer is needed, as opposed to the form factor method. 
This method, however, requires a large number of configurations to reduce statistical errors.
Performing calculations at four different quark masses, we discuss a quark mass dependence of the NEDM 
in full QCD. 
In particular it is interesting to see whether the NEDM vanishes or not at zero quark mass 
as theoretically predicted.
There exists a previous study of the NEDM in lattice QCD with 2-flavor dynamical domain-wall quarks 
using the form factor method \cite{RBC}. Unfortunately the signal of the NEDM is consistent 
with zero within a large error, so that only an upper bound is obtained for a value of $d_N^{(1)}$.

The organization of this paper is as follows.
In Sec.\ref{sec:2}, we  give a definition of NEDM in the presence of uniform and static
electric field. We then consider how this external electric field is introduced on the lattice,
and discuss a violation of uniformity of the electric field due to boundaries of the finite lattice.
We also explain our method of extracting the NEDM from the nucleon propagator.
In Sec.\ref{sec:3}, simulation parameters for dynamical configurations are briefly summarized. 
Our main results are given in Sec.\ref{sec:4}.
Summary and discussions of this paper are presented in Sec.\ref{sec:5}.
In this paper we set $a=1$ unless necessary.

\section{Method of the NEDM calculation}\label{sec:2}
\subsection{Definition of the NEDM}
In the presence of the constant and uniform electric field $\vec E$,
a change of energy for the nucleon state due to the $\theta$ term is denoted as
\begin{equation}
  \Delta \mathcal E_{\rm CP} = d_N^{(1)}\theta\vec S\cdot\vec E + \mathcal O(\vec E^3\theta,\vec E\theta^3 )
\end{equation}
with the nucleon spin vector $\vec S$. 
This leads to the following extraction of $d_N^{(1)}$ for $\vec E =(0,0,E_z)$:
\begin{equation}
  \mathcal E^\theta_{+}(E_z) - \mathcal E^\theta_{-}(E_z)
  = d_N^{(1)}\theta E_z+ \mathcal O(E_z^3\theta,E_z\theta^3 )
\label{eq:EnergyDiff}
\end{equation}
where $\mathcal E_{\pm}^\theta(E_z)$ is an energy of the nucleon state with $S_z=\pm 1/2$ 
in the presence of the electric field $\vec E =(0,0,E_z)$.
Hereafter we simply denote  $d_N^{(1)}$ as $d_N$.

\subsection{Introduction of the electric field on the lattice and a boundary effect}
On the lattice,
an external electric field $E_k$ is introduced into link variables via a replacement
in the Wilson-Dirac operator with
\begin{equation}
  U_k(x)     \longrightarrow e^{e_q E_k t}U_k(x)       \equiv U^q_k(E,x),\quad
  U^\dag_k(x) \longrightarrow e^{-e_q E_k t}U^{\dag}_k(x)\equiv \overline U^{q}_k(E,x), 
\end{equation}
where $e_q$ is an electric charge of a quark flavor $q$.
Note here that a complex factor $i$ does not appears in the exponent since $E_k$ 
is defined in the "Minkowski" space while $t$ is the time coordinate of
the "Euclidean" lattice, and 
therefore $U_k^q(E,x)$ is no more unitary
\footnote{ 
With the electric field in the Minkovski space, however, the energy
difference \ref{eq:EnergyDiff} becomes real in the lattice calculation,
so that it can be extracted from the ratio nucleon propagators between
spin-up and spin-down\cite{Aoki}.}.
In our calculation we have made this replacement of the link variables
only for 
the valence quark \cite{Shintani_wE}, while gauge configurations have been already generated 
without this replacement in the quark determinant. This "approximation" is equivalent to ignoring 
the "disconnected" contribution from the electric field.  It is noted that, at the first order 
of the electric field, this contribution vanishes for the 3 flavor QCD with $m_u=m_d=m_s$, 
since the disconnected contribution does note depend on the quark flavor in the flavor SU(3) limit, 
and therefore is proportional to $e_u+e_d+e_s=0$ after summing over 3 flavors.

As discussed in Ref.\cite{Shintani_wE}, 
the introduction of the "Minkowski" electric filed  
destroys the periodic boundary condition of link variable $U^q_k(E,x)$, 
so that translational invariance of the electric field is violated at the temporal boundary.
Indeed an effective electric fields, defined by 
$E_k(t_E)=\{\ln U_k^q(E, t_E+1)-\ln U_k^q(E,t_E)\}/e_q$ with $U_k(t_E)=1$, becomes
\begin{equation}
  E_k(t_E) = \left\{
  \begin{array}{cl}
     E_k       & \textrm{at }t_E=1,2,\cdots T-1 \\
     -(T-1) E_k& \textrm{at }t_E=T
  \end{array}
  \right. ,
\end{equation}
where $t_E$ runs from 1 to $T$, so that $t_E = T+1$ is equal to $t_E=1$.
A strong anti-electric field is generated between $t_E = 1$ and $t_E = T$ 
in order to cancel the constant electric field $E_k$, so that $\sum_{t_E=1}^T E_k(t_E)$ vanishes.
We denote the place where the large gap exist as $t_{\rm Gap}$,  and $t_{\rm Gap}=0$ in the above case.
Since the definition of the NEDM in (\ref{eq:EnergyDiff}) requires the constant electric field, 
the NEDM should only be measured far away form the boundary to avoid an effect of the anti-electric field. 
In this work we keep distances between source/sink points and the boundary as large as possible.
In particular we fix $|t_{\rm Gap} - t_{\rm src} + 1| = T/2$, 
where $t_{\rm src}$ is the time coordinate of the source point.

In the previous study \cite{Shintani_wE} we found that  
the good sampling of the topological charge is important for obtaining
the reliable signal of NEDM. 
At least an order of a few thousands configurations was needed to 
realize the symmetric and Gaussian  
distribution in quenched QCD.
On the other hand, 
a number of full QCD configurations in this study is limited to  $700\sim 750$ at each quark mass.
We therefore use several source points for each configuration to increase statistics.
In accordance with the change of the source point, 
we shift the boundary point, keeping the distance between the source point 
and the boundary as large as possible,
to avoid the boundary effect mentioned above.

\subsection{Spinor structure of the nucleon propagator and EDM}
In the presence of the electric field $\vec E$ and the CP violating $\theta$ term, 
the explicit form of the nucleon propagator becomes
\begin{eqnarray}
  \langle N_{\alpha} \bar N_{\beta} \rangle_{\theta}(\vec E,t)
 &=& Z_N(E^2)
    \Big[\left(1 + A_N(E^2)\theta\vec\sigma\cdot\vec E\right)\nonumber\\
 &\times& \exp\big(-\mathcal E_N(E^2)t
  - \frac{d_N\theta}{2}\vec{\sigma}\cdot\vec{E}t\big) \Big]_{\alpha\beta} + \cdots ,
\label{eq:NN_with_E}
\end{eqnarray}
with an overall amplitude $Z_N(E^2)$, a coefficient $A_N(E^2)$ and 
an energy $\mathcal E_N(E^2)$\footnote{Note that because of the acceleration for a charged particle 
the energy of proton increases as a time increases \cite{Detmold}.
However this contribution is canceled out in the following ratio.}.
The ellipse represents $\mathcal O(E^3\theta,\theta^2)$  terms and 
contributions from excited states.
Here $\langle \mathcal O \rangle_\theta$ represents a vacuum expectation
value in reweighting method as 
\begin{eqnarray}
Z_\theta \langle \mathcal O\rangle_\theta = 
Z_{\theta=0}\langle \mathcal O e^{i\theta Q}\rangle_{\theta=0} 
\end{eqnarray}
with topological charge $Q$, which is evaluated by the cooling method
in our calculation\cite{Shintani_form, Shintani_wE}.

In order to extract $d_N$ from the nucleon propagator in the above equation,
we consider the following ratio between different spinor components;
\begin{eqnarray}
  R_3(E,t;\theta)
  &=& \frac{R^{\rm naive}_3(E,t;\theta)}{R^{\rm naive}_3(E=0,t;\theta)}
      \frac{R^{\rm naive}_3(E=0,t;\theta=0)}{R^{\rm naive}_3(E,t;\theta=0)} \nonumber\\
  &\simeq& \frac{1+\theta A_N(E^2) E}{1-\theta A_N(E^2) E}\exp[-d_N\theta Et], 
  \label{eq:R_def}
\end{eqnarray}
where
\begin{eqnarray}
  R_3^{\rm naive}(E,t;\theta)
  &=& \frac{\langle N_1 \bar N_1 \rangle_{\theta}((0,0,E),t)}
           {\langle N_2 \bar N_2 \rangle_{\theta}((0,0,E),t)}
  = \frac{1+A_N\theta(E^2) E}{1-A_N\theta(E^2) E}
    \exp[ -d_N\theta Et ] + \cdots,
\end{eqnarray}
Three additional $R_3^{\rm naive}$'s in $R_3(E,t;\theta)$ are introduced to remove contamination 
coming from $\theta=0$ and/or $E=0$ terms due to the insufficient statistics. 
In addition, to remove fictitious $E^{2n}\theta$ contribution, we construct the ratio
\begin{eqnarray}
  R_3^{\rm corr}(E,t;\theta)
&=& \frac{R_3(E,t;\theta)}{R_3(-E,t;\theta)}
 =  \frac{R_3^{\rm naive}(E,t;\theta)}{R_3^{\rm naive}(-E,t;\theta)}
    \frac{R_3^{\rm naive}(-E,t;\theta=0)}{R_3^{\rm naive}(E,t;\theta=0)}\nonumber\\
&\simeq& \left(\frac{1+\theta A_N^1(E^2)E}{1-\theta A_N^1(E^2)E}\right)^2\exp[-2d_N\theta E t],
\label{eq:R_corr_def}
\end{eqnarray}
with $R_3(E,t;\theta)$ in eq.(\ref{eq:R_def}).
For later use we define the "effective $d_N$"  as
\begin{equation}
  2d_N\theta E
  = \ln\left[\frac{R^{\rm corr}_3(E,t-1;\theta)}
                  {R^{\rm corr}_3(E,t;\theta)}\right],
  \label{eq:eff_R}
\end{equation}
which is validated in large $t$ where 
the nucleon asymptotic state dominates,

\section{Simulation parameters}\label{sec:3}
In our calculation, we employ 2-flavor dynamical QCD configurations, generated
by CP-PACS collaboration \cite{cppacs} with
the RG-improved (Iwasaki) gauge action and the clover quark action on $24^3\times 48$ lattice at $\beta=2.1$. The corresponding lattice spacing, determined by the rho meson mass $m_\rho = 768.4$ MeV, is $a^{-1}\simeq 1.8$ GeV ($a\simeq 0.11$ fm).

The valence quark mass is chosen to be same with the sea quark mass.
The pion mass $m_{PS}$ becomes 1.13, 0.93, 0.76 and 0.53 GeV
at $K_{\rm sea}=0.1357,\,0.1367,\,0.1374$ and $\,0.1382$, respectively.
Table \ref{tab:lat_param} lists lattice parameters used in our simulation.
Throughout this paper statistical errors are estimated by the jack-knife method,
whose bin size is 5 configurations, equivalent to 25 HMC trajectories. 
We employ a local sink and a smeared source for all three quark propagators in
the nucleon 2-pt function, 
with the exponential smearing, $f(r)=Ae^{-Br}$.
Parameters $(A,B)$ depend on the quark mass as shown in Table \ref{tab:lat_param}.

We take 16 different source points separated by 3 lattice units in temporal direction and
maximally separated in spatial directions.
Averaging these source points, a total number of statistics is 
more than 10,000. 

Throughout our study, the value of electric field and $\theta$ are fixed to
$E=0.004$ and $\theta=0.025$, 
which are the most suitable choice to reduce statistical errors \cite{Shintani_wE}.

\section{Results}\label{sec:4}
\subsection{Topological charge distribution and nucleon mass}
\begin{figure}[h]
\begin{center}
\vskip 10mm
\includegraphics[width=140mm, angle=0] {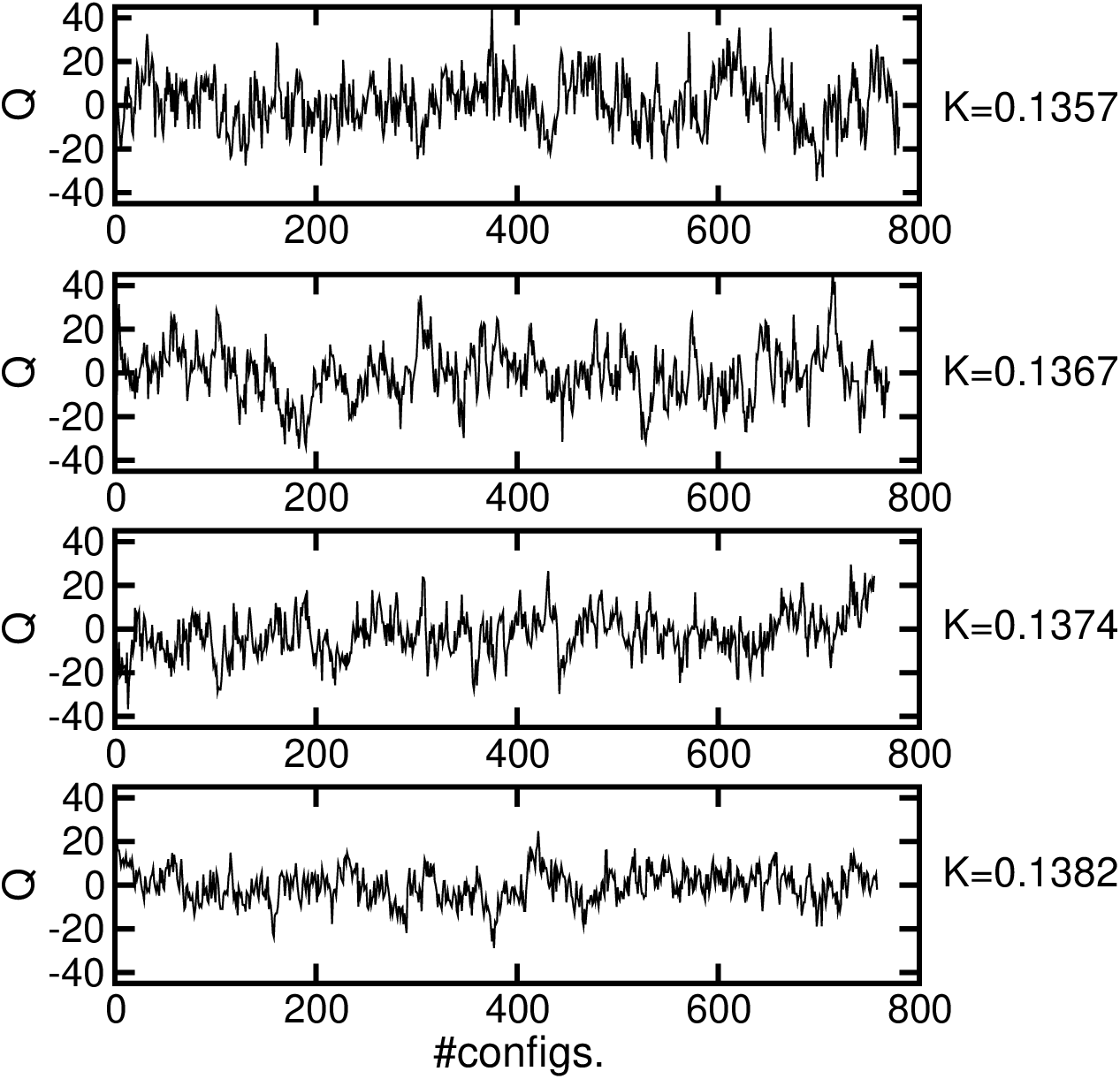}
\caption{Time histories of the topological charge at each sea quark mass.}
\label{fig:top}
\end{center}
\end{figure}

\begin{figure}[h]
\begin{center}
\vskip 10mm
\includegraphics[width=75mm, angle=0] {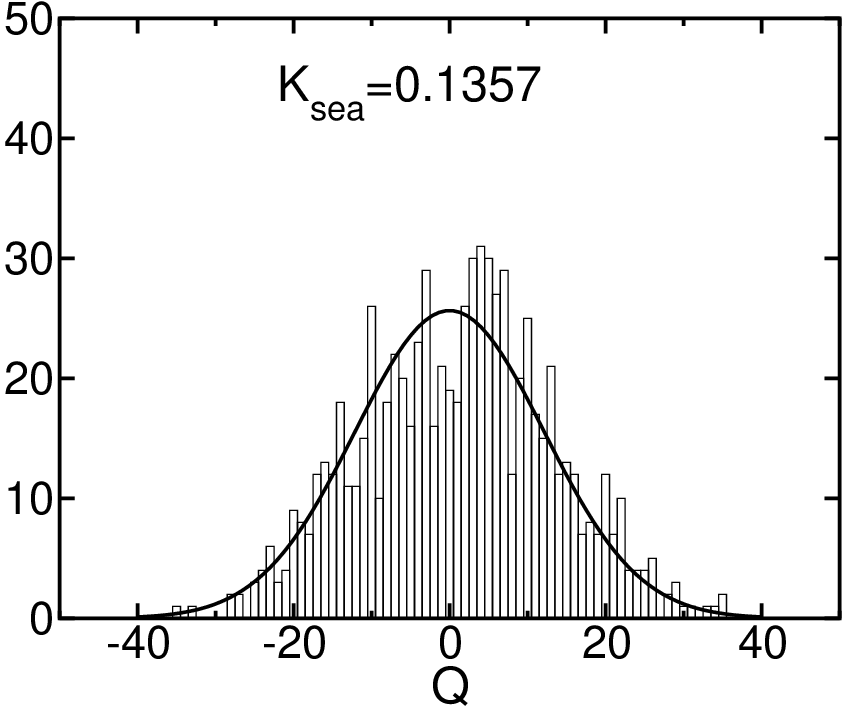}
\hspace{5mm}
\includegraphics[width=75mm, angle=0] {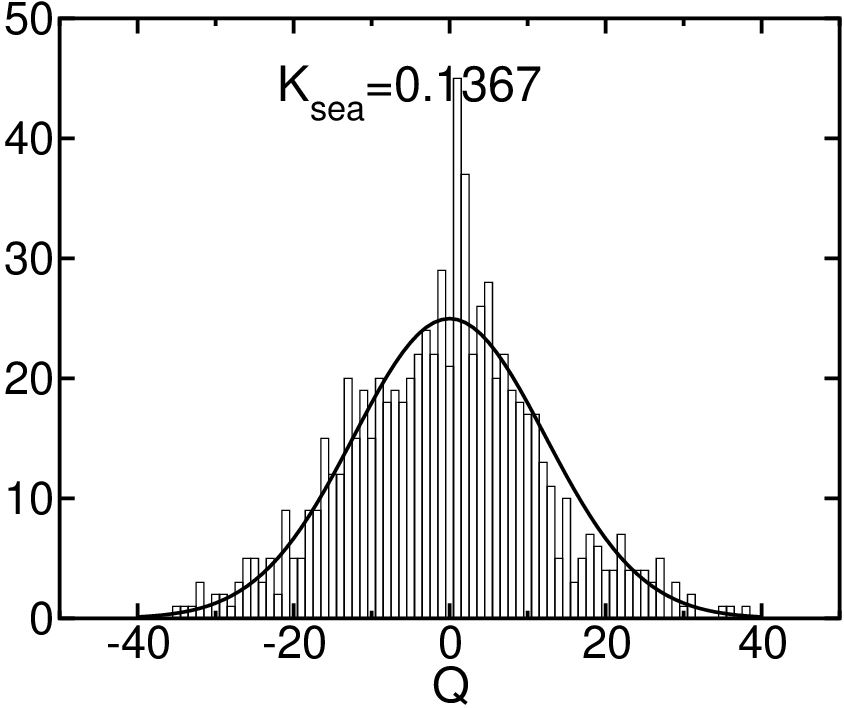}
\vskip 5mm
\includegraphics[width=75mm, angle=0] {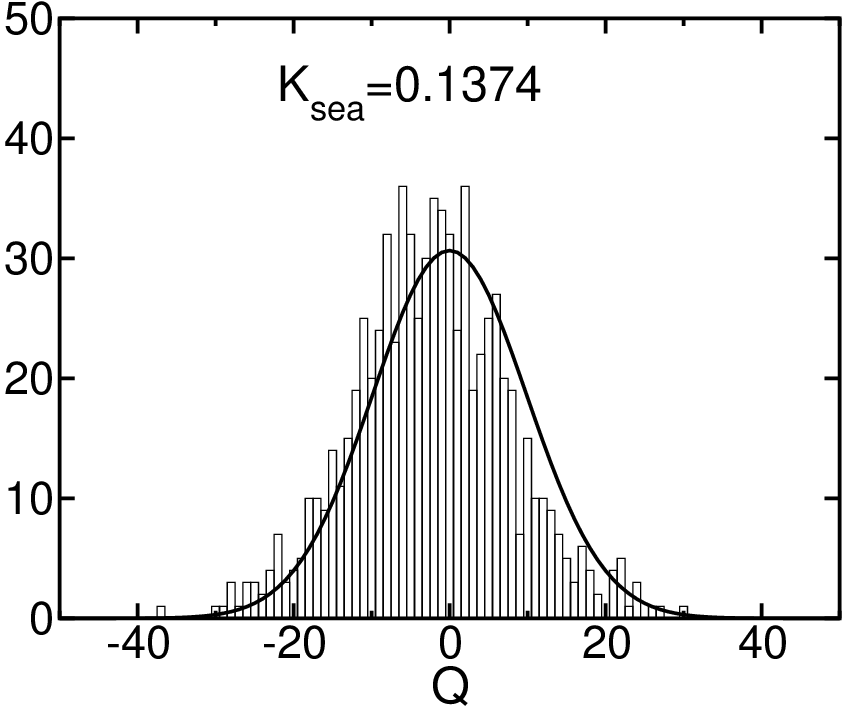}
\hspace{5mm}
\includegraphics[width=75mm, angle=0] {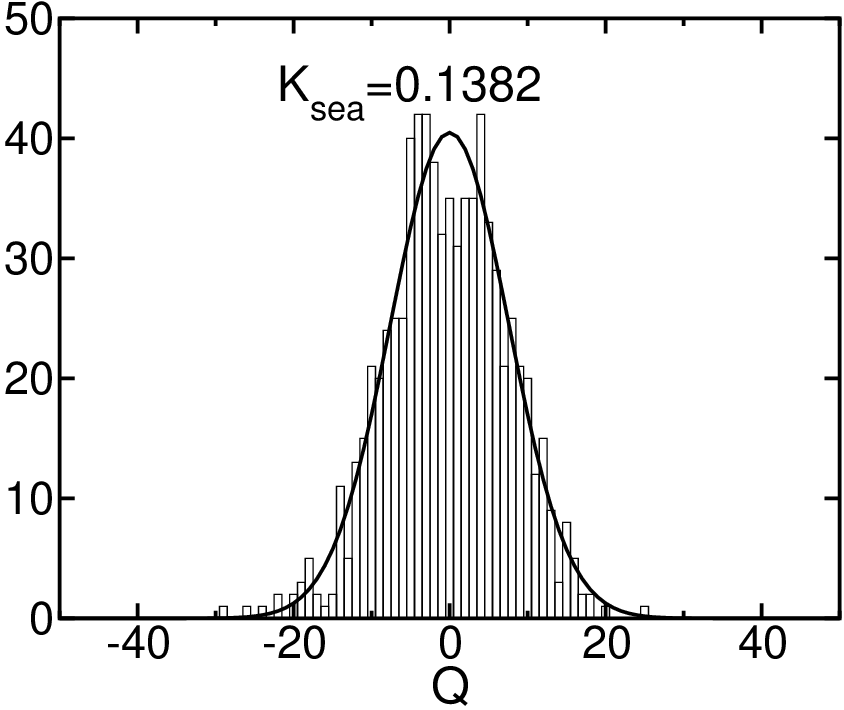}
\caption{Histograms of the topological charge at each sea quark mass. 
The solid line denotes the Gaussian distribution determined from  
$\langle Q\rangle$ and $\langle Q^2\rangle$.}
\label{fig:hist}
\end{center}
\end{figure}

We measure the topological charge of each configuration after 50 cooling steps,
using the $\mathcal O(a^2)$ improved definition of the topological charge density given by
\begin{equation}
  Q_{\rm improved} = \frac{1}{32\pi^2}\varepsilon_{\mu\nu\alpha\beta}
  \left( a_0{\rm Tr}[L^{1\times 1}_{\mu\nu}L^{1\times 1}_{\alpha\beta}]
      + 2a_1{\rm Tr}[L^{1\times 2}_{\mu\nu}L^{1\times 2}_{\alpha\beta}] \right)
  \label{eq:Q_def}
\end{equation}
with $a_0=5/3$ and $a_1=-1/12$ \cite{impQ,cppacs_Q},
where $L^{1\times 1}_{\mu\nu}$ and $L^{1\times 2}_{\mu\nu}$ are $1\times 1$ and $1\times 2$ Wilson loops, respectively.
We show the time history of the topological charge in Fig.\ref{fig:top} and 
its histogram in Fig.\ref{fig:hist} at four quark masses. 
The distributions of the topological charge is more or less gaussian at all quark masses.
As the sea quark mass decreases, the width of the gaussian distribution 
becomes narrower in accordance with the theoretical expectation.  

\begin{figure}[h]
\begin{center}
\vskip 10mm
\includegraphics[width=140mm, angle=0] {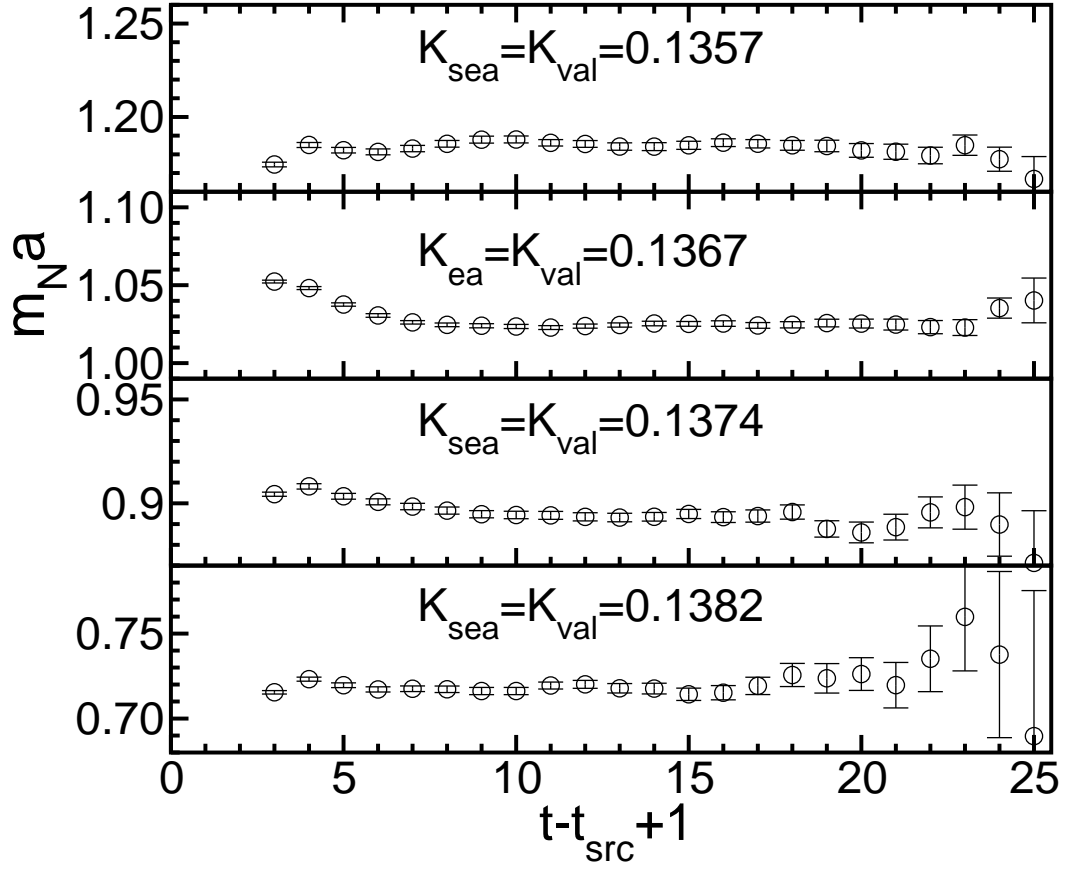}
\caption{The nucleon effective mass as a function of time in lattice unit  without electric field.
 The average over 16 source sets on each configuration is taken.}
\label{fig:Nmass}
\end{center}
\newpage
\end{figure}

The effective mass of the nucleon at four quark masses is plotted in Fig.\ref{fig:Nmass}. 
The average  over 16 sources gives enough statistics to produce a clear plateau at all cases. 
We see that the plateau for the nucleon state starts at $t-t_{\rm src}+1 = 6$.
The global fit of the propagator from $t-t_{\rm src}+1=$ $9$ to $15$ gives a value 
of the nucleon mass with a very small error, as shown in Table \ref{tab:lat_param}.

\subsection{Signal of EDM}
\begin{figure}[h]
\begin{center}
\vskip 10mm
\includegraphics[width=75mm, angle=0] {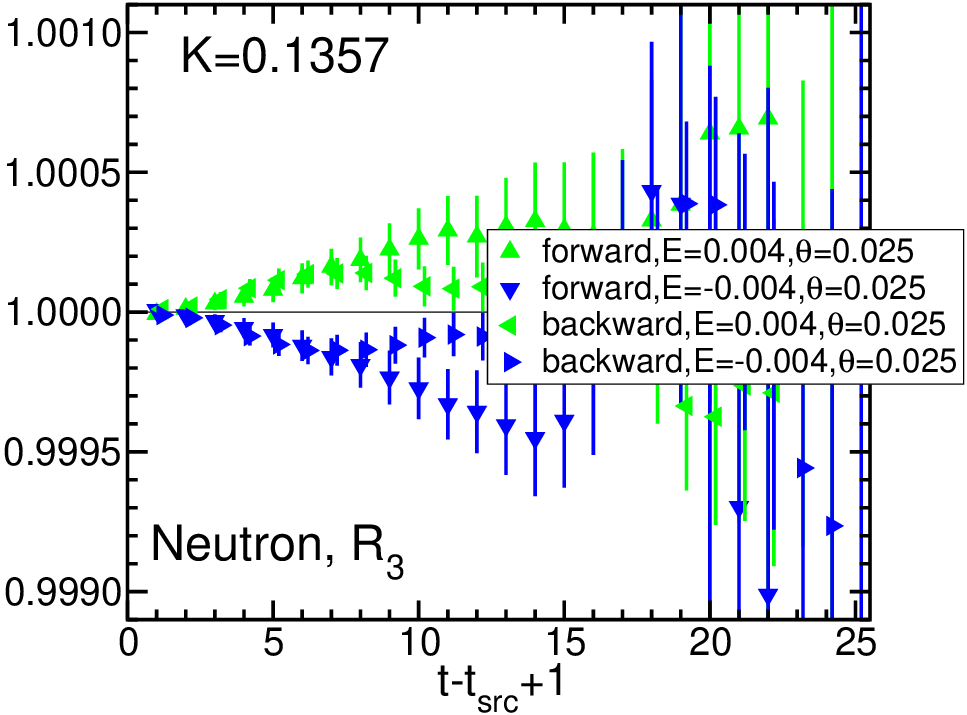}
\hspace{5mm}
\includegraphics[width=75mm, angle=0] {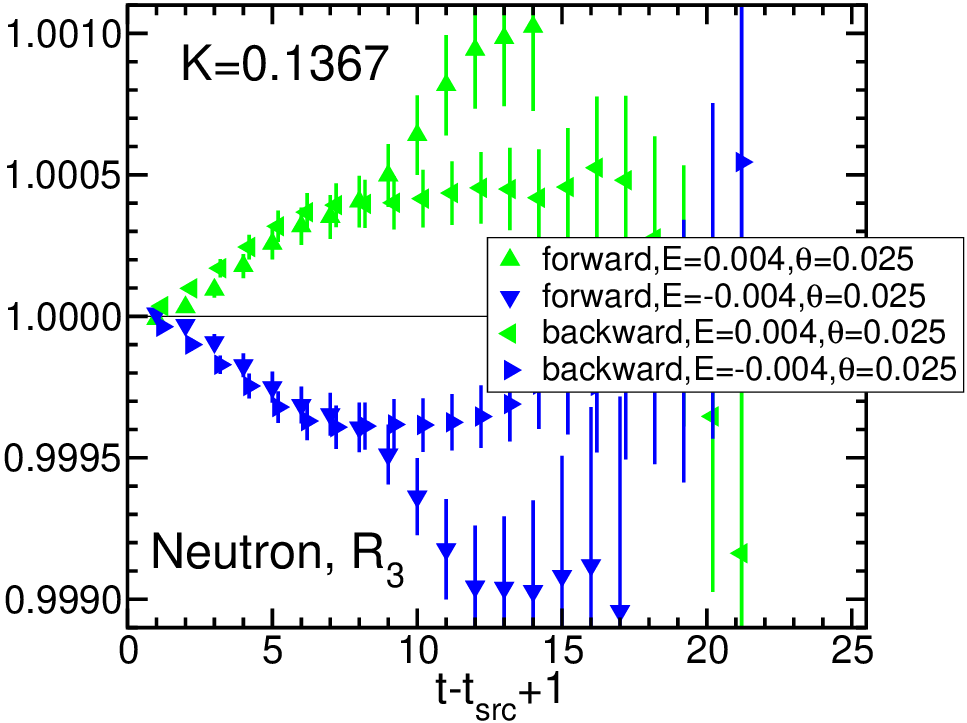}
\vskip 10mm
\includegraphics[width=75mm, angle=0] {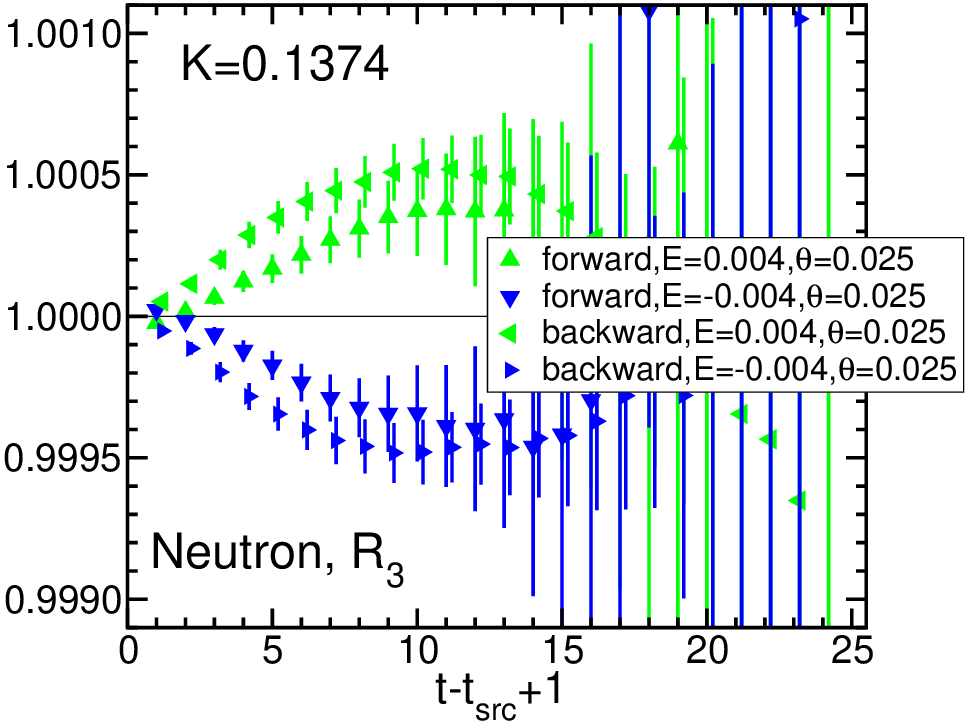}
\hspace{5mm}
\includegraphics[width=75mm, angle=0] {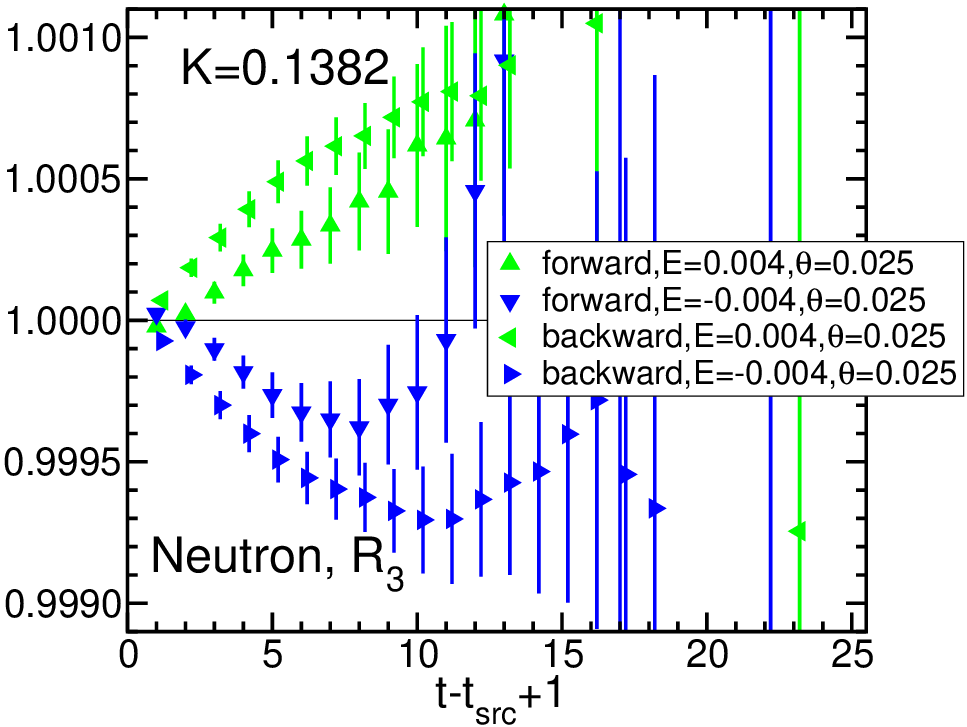}
\caption{Comparison between forward and backward propagation of $R_3$ for neutron 
as a function of time at $E=\pm0.004$ and $\theta=0.025$.
The average over 16 source sets ($t_{\rm src}=3,6,\cdots,48$) is taken.}
\label{fig:R_rev}
\end{center}
\newpage
\end{figure}

In the previous study we found that the influence due to the gap of the electric field on the EDM signal disappears
at $|t_{\rm Gap}-t_{\rm src}+1| \ge 5$ at $a=0.1$ fm in quenched QCD \cite{Shintani_wE}.
Since we fix $|t_{\rm Gap}-t_{\rm src}+1|=T/2 = 24$ in this study, the above condition is well satisfied.
In this set up, the backward propagation of the nucleon becomes identical to the forward propagation 
in the infinite statistics, so that we can take an average over both to increase statistics.
In Fig.\ref{fig:R_rev}, we compare the time dependence of $R_3(E,t;\theta)$ between 
forward and backward propagation for neutron and proton. 
We have found that two results marginally agree with each other at $t-t_{\rm src}+1\le 10$.
Here we consider that the difference at $t-t_{\rm src}+1>10$ is caused by the statistical noises, 
and thus the signal of EDM is obtained only at $t-t_{\rm src}+1\le 10$. 
Therefore a fitting range should be chosen as $6 \le t-t_{\rm src}+1\le 10$.
Since $T - (t - t_{\rm Gap}) = T/2 - 9 = 15 $, the distance between the gap and the end-point of the fitting ranged at $t -t_{\rm src} +1 = 10$, is  
much larger than $5$, the gap of the electric field does not affect the EDM signal.
 
\begin{figure}[h]
\begin{center}
\vskip 10mm
\includegraphics[width=85mm, angle=0] {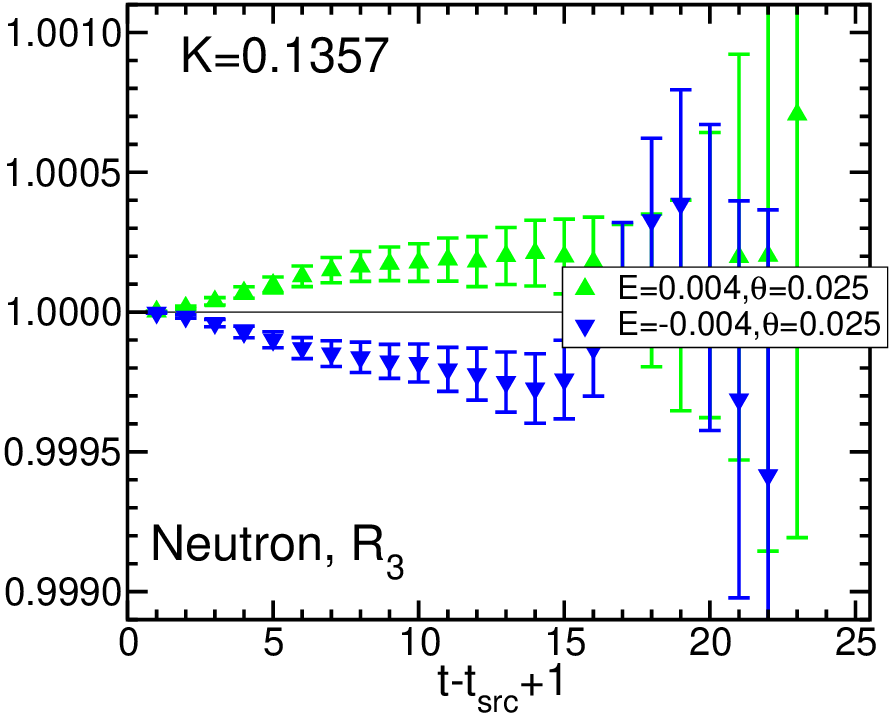}
\vskip 10mm
\includegraphics[width=85mm, angle=0] {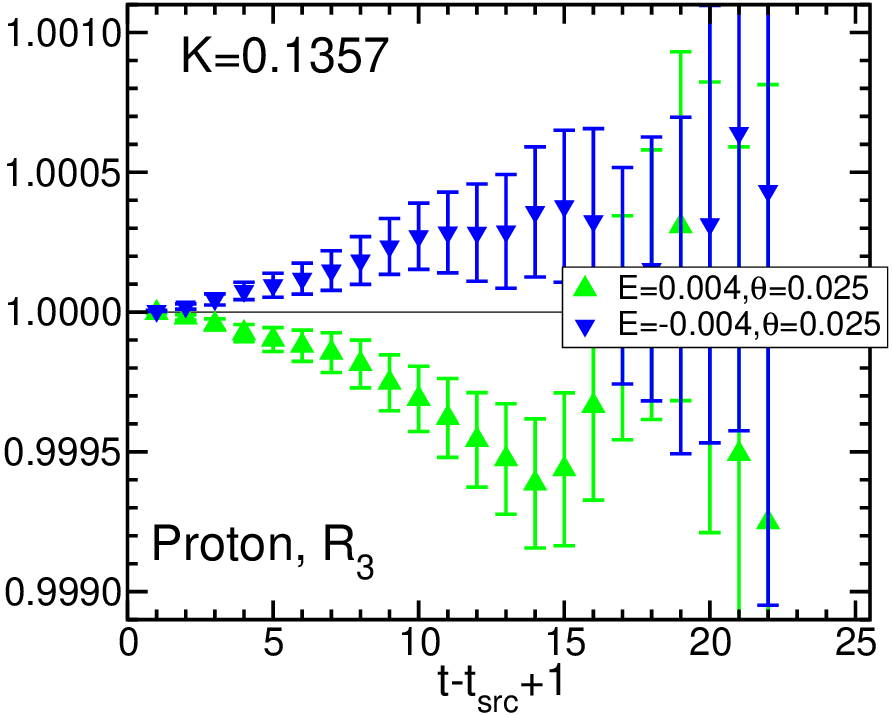}
\caption{$R_3$ as a function of time at $K=0.1357$, 
after averaging 16 source sets ($t_{\rm src}=3,6,\cdots,48$),
with $E=\pm0.004$ and $\theta=0.025$ for neutron (top) and proton (bottom).
The different symbols denote the different signs of the electric field.}
\label{fig:R_K01357}
\end{center}
\end{figure}

\begin{figure}[h]
\begin{center}
\vskip 10mm
\includegraphics[width=85mm, angle=0] {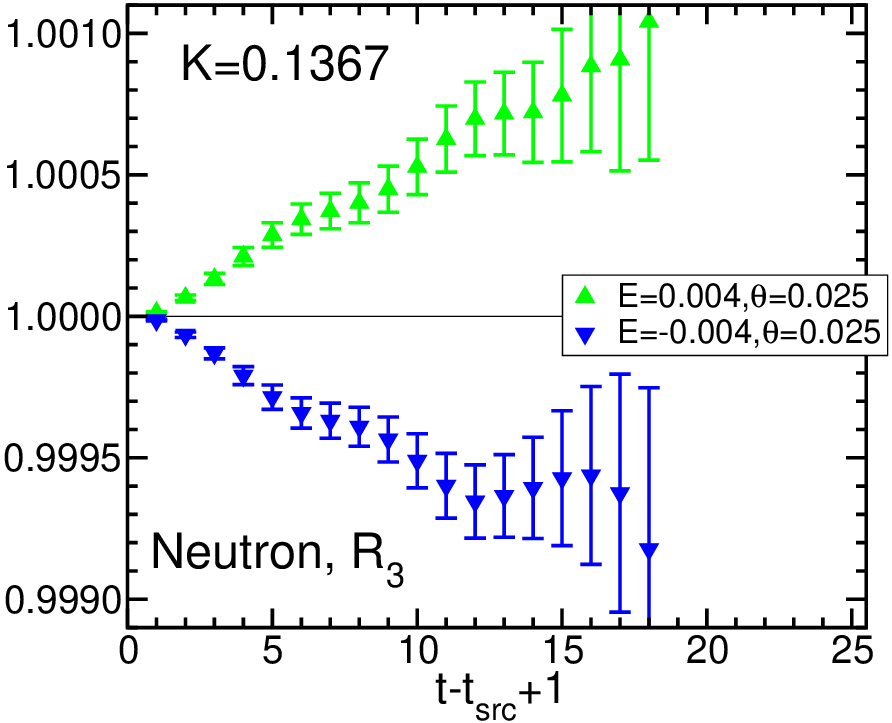}
\vskip 10mm
\includegraphics[width=85mm, angle=0] {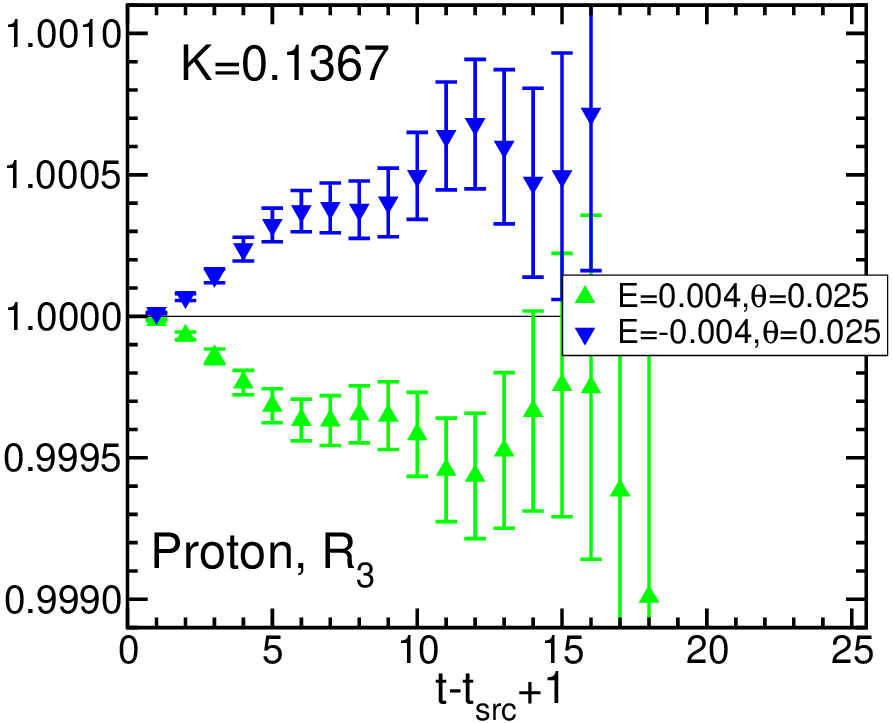}
\caption{Same as Fig. \ref{fig:R_K01357} at $K=0.1367$.}
\label{fig:R_K01367}
\end{center}
\end{figure}

\begin{figure}[h]
\begin{center}
\vskip 10mm
\includegraphics[width=85mm, angle=0] {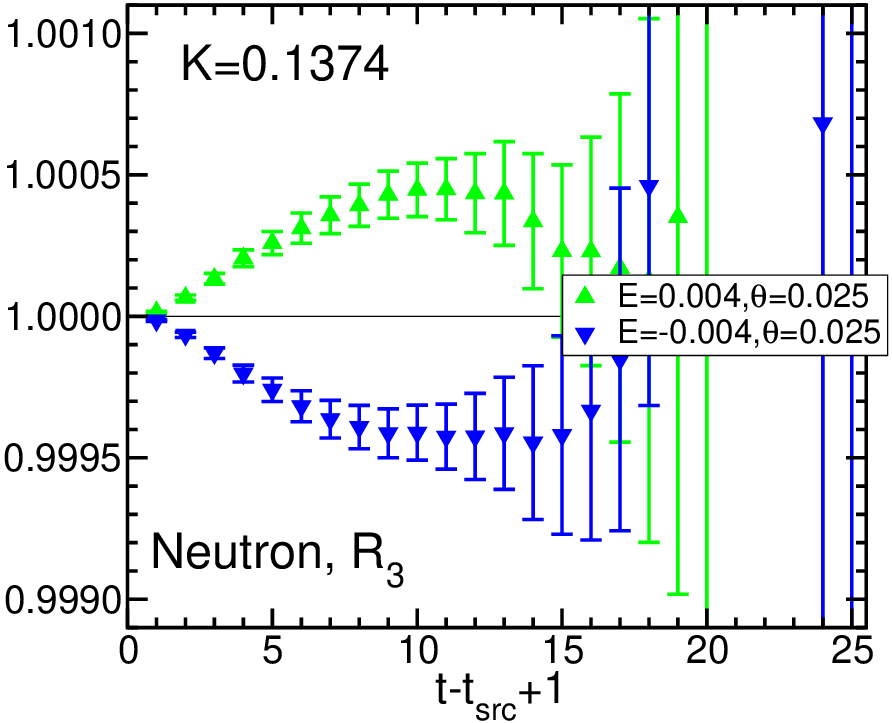}
\vskip 10mm
\includegraphics[width=85mm, angle=0] {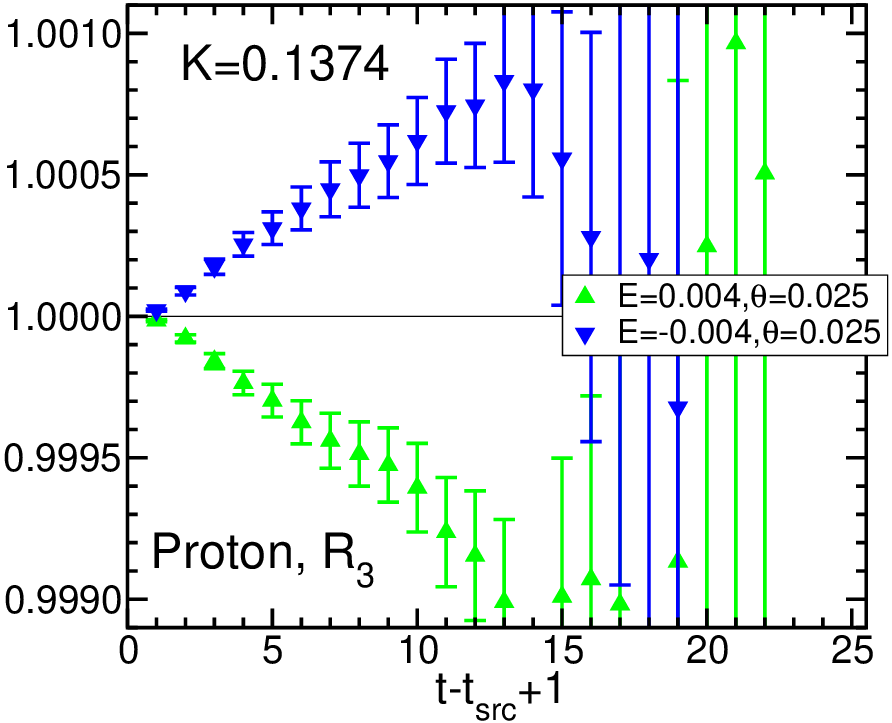}
\caption{Same as Fig. \ref{fig:R_K01357} at $K=0.1374$.}
\label{fig:R_K01374}
\end{center}
\end{figure}

\begin{figure}[h]
\begin{center}
\vskip 10mm
\includegraphics[width=85mm, angle=0] {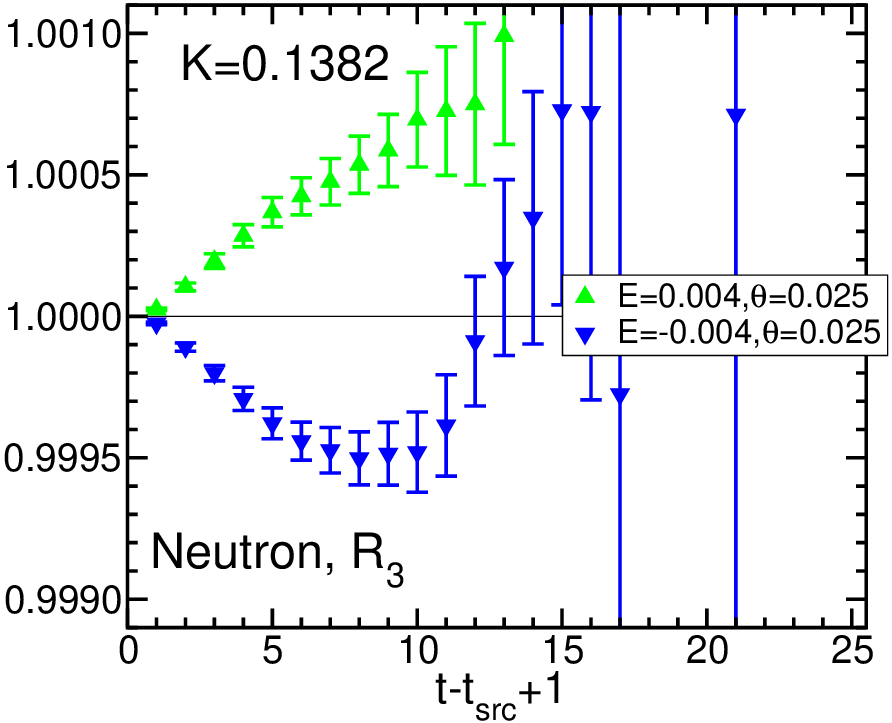}
\vskip 10mm
\includegraphics[width=85mm, angle=0] {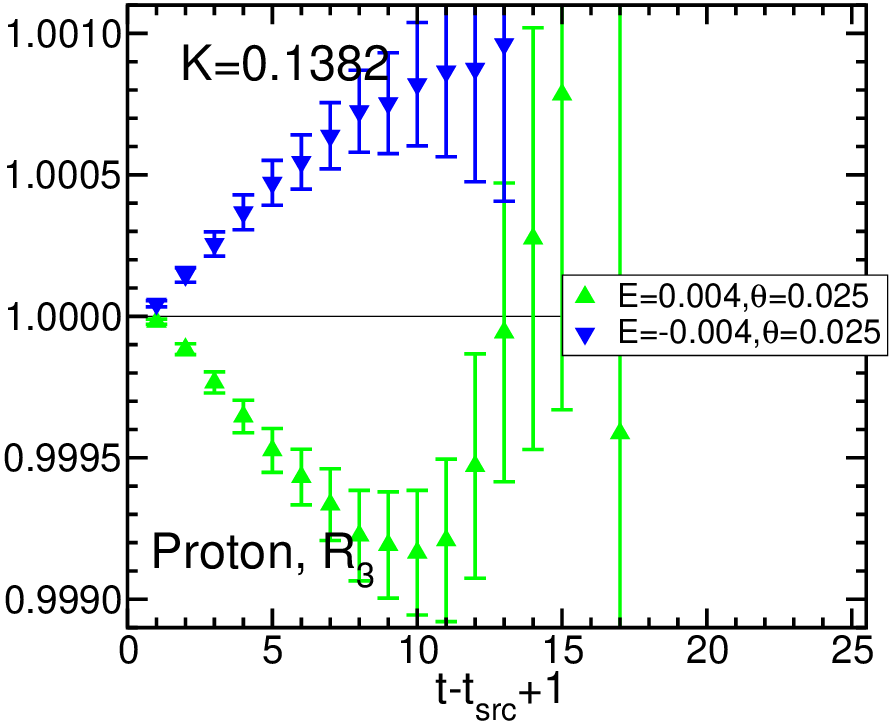}
\caption{Same  as Fig. \ref{fig:R_K01357} at $K=0.1382$.}
\label{fig:R_K01382}
\end{center}
\end{figure}

\begin{figure}[h]
\begin{center}
\vskip 10mm
\includegraphics[width=85mm, angle=0] {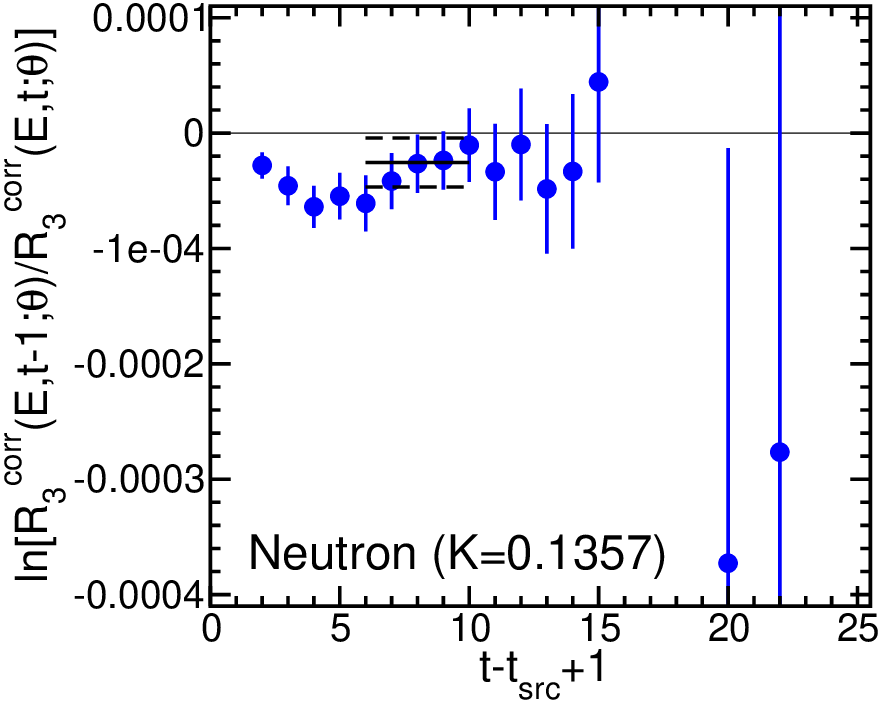}
\vskip 10mm
\includegraphics[width=85mm, angle=0] {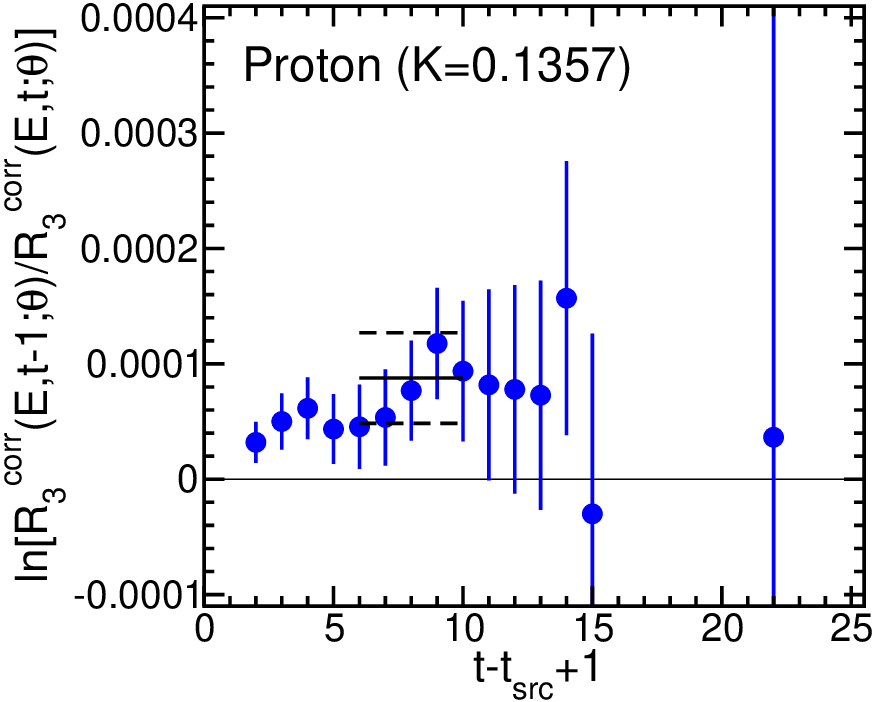}
\caption{The effective mass plot of $R_3^{\rm corr}$ defined in eq.(\ref{eq:eff_R})
as a function of time in lattice unit at $K=0.1357$ for neutron (top) and proton (bottom). 
The solid line denotes the central value of the fitting result and two dash lines indicates 
an error band.}
\label{fig:Reff_K01357}
\end{center}
\end{figure}

\begin{figure}[h]
\begin{center}
\vskip 10mm
\includegraphics[width=85mm, angle=0] {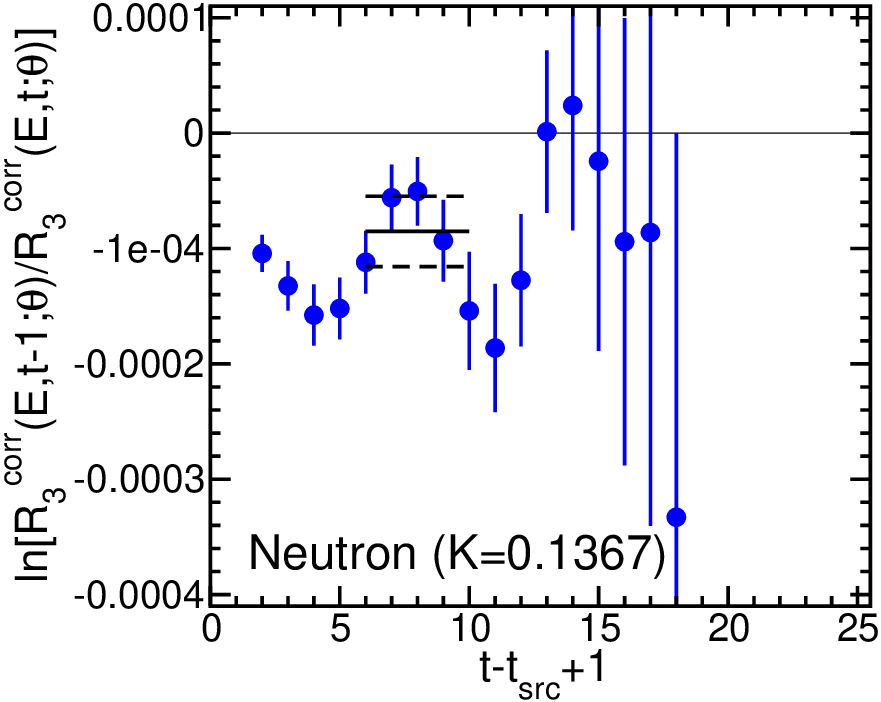}
\vskip 10mm
\includegraphics[width=85mm, angle=0] {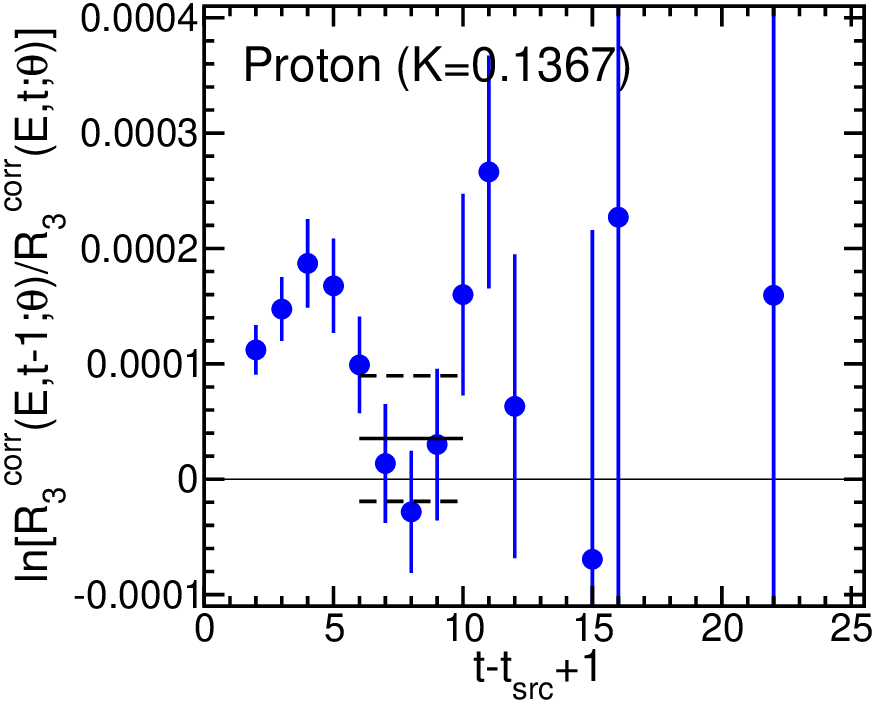}
\caption{Same  as Fig. \ref{fig:Reff_K01357} at $K=0.1367$.}
\label{fig:Reff_K01367}
\end{center}
\end{figure}

\begin{figure}[h]
\begin{center}
\vskip 10mm
\includegraphics[width=85mm, angle=0] {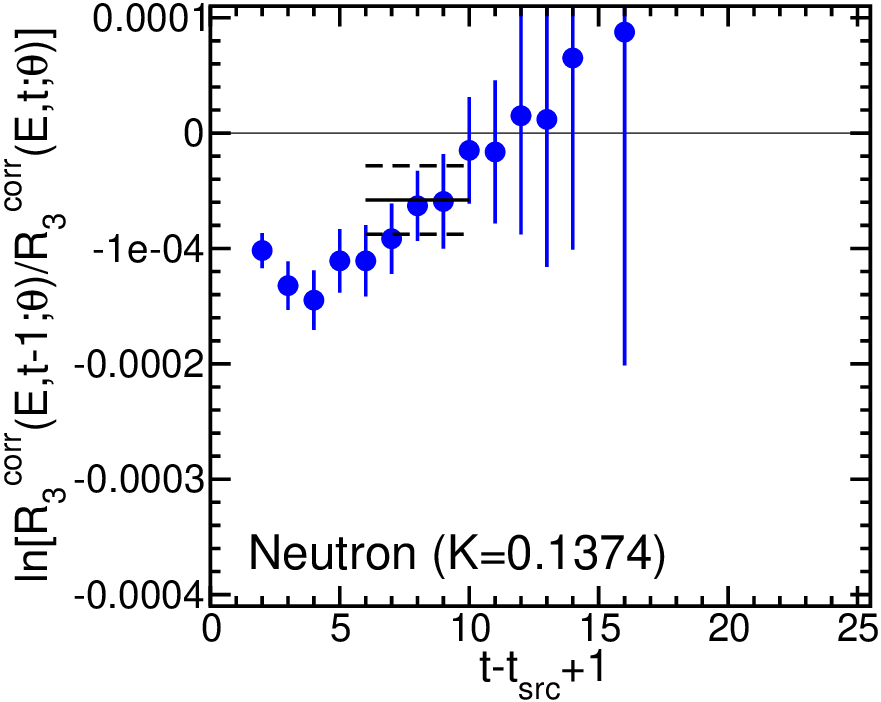}
\vskip 10mm
\includegraphics[width=85mm, angle=0] {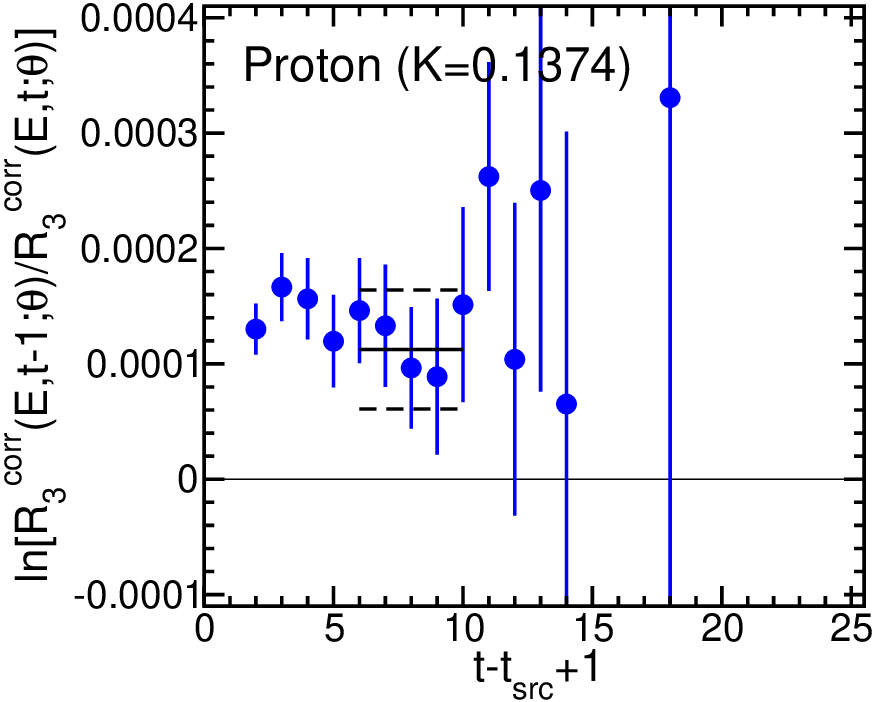}
\caption{Same as Fig. \ref{fig:Reff_K01357} at $K=0.1374$.}
\label{fig:Reff_K01374}
\end{center}
\end{figure}

\begin{figure}[h]
\begin{center}
\vskip 10mm
\includegraphics[width=85mm, angle=0] {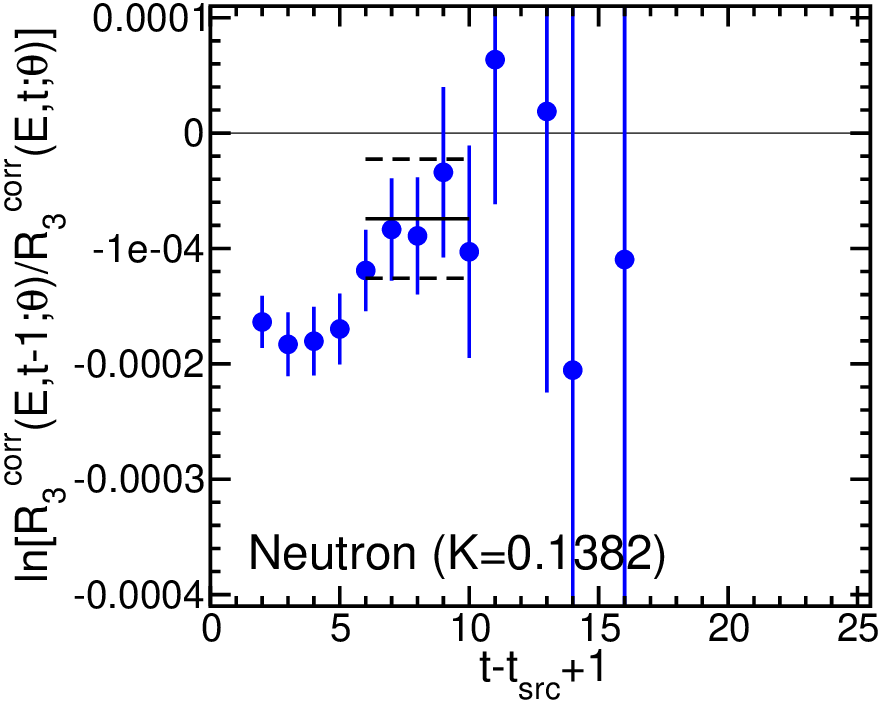}
\vskip 10mm
\includegraphics[width=85mm, angle=0] {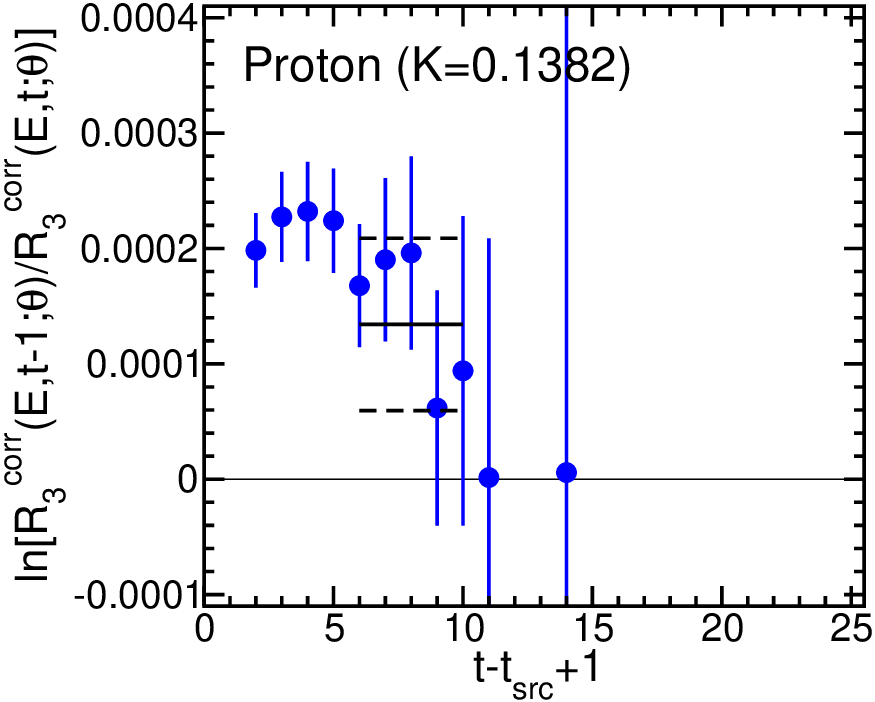}
\caption{Same as Fig. \ref{fig:Reff_K01357} at $K=0.1382$.}
\label{fig:Reff_K01382}
\end{center}
\end{figure}

In Fig.\ref{fig:R_K01357}-\ref{fig:R_K01382}, the time dependence of $R_3(E,t;\theta)$ 
with $E=\pm 0.004$ is plotted for the neutron and proton at each quark mass. 
At all quark masses, the signal for non-zero EDM can be seen at $6 \le t-t_{\rm src}+1\le 10$, 
and the signal changes its sign as $E$ does.
To estimate the size of nucleon EDM, we consider an effective mass of 
$R_3^{\rm corr}(E,t;\theta)$ defined in eq.(\ref{eq:eff_R}), and plot it  
in Fig.\ref{fig:Reff_K01357}-\ref{fig:Reff_K01382}.  
Although errors and fluctuations are large, non-zero signals for both proton and neutron EDM 
have been  observed in full QCD simulations for the first time. 
By fitting $R_3^{\rm corr}(E,t;\theta)$ with an exponential function in eq.(\ref{eq:R_corr_def})
at $6 \le t-t_{\rm src}+1\le 10$,
we obtain the value of $d_N^{(1)}$, which is given in Table \ref{tab:EDM}. 
Signs of EDM are opposite between proton and neutron. This agrees with 
the quenched result \cite{Shintani_wE} and with the CHPT prediction \cite{CHPT_dn}.

\subsection{Mass dependence of the nucleon EDM}
\begin{figure}[h]
\begin{center}
\vskip 10mm
\includegraphics[width=85mm, angle=0] {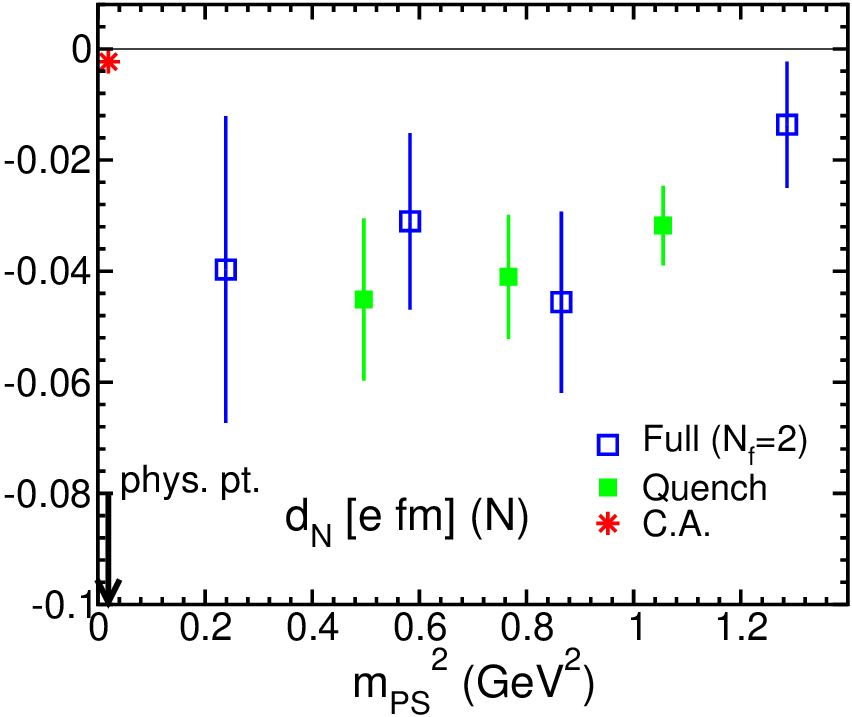}
\vskip 10mm
\includegraphics[width=85mm, angle=0] {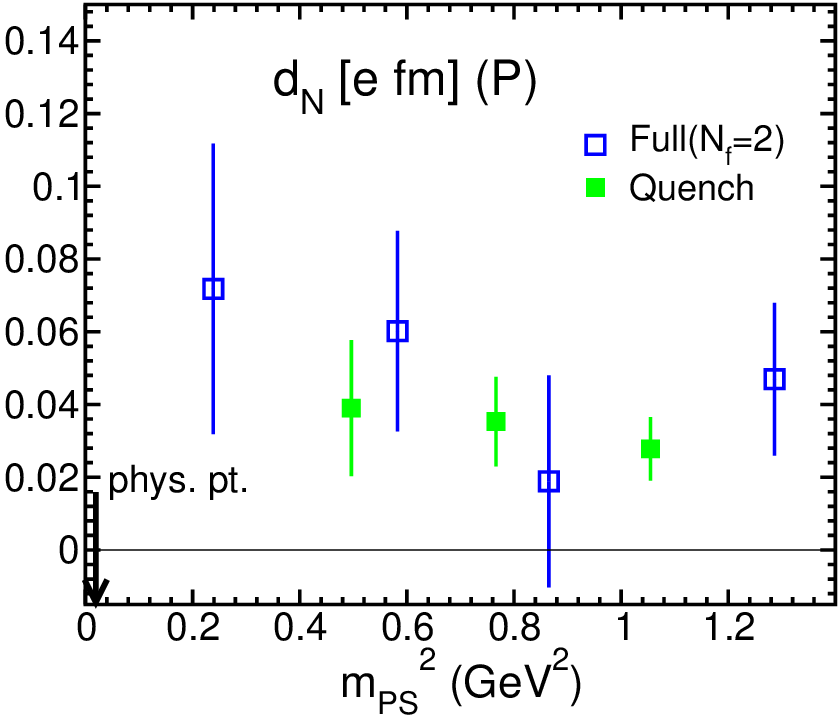}
\caption{EDM as a function of the pseudoscalar meson mass squared $m_{PS}^2$ 
for neutron (top) and proton (bottom).
The arrow shows the physical point of the pion mass squared, $m_\pi^2=0.0195$ GeV${}^2$, and 
the star symbol denotes the result of the current algebra \cite{model_CA}.} 
\label{fig:dN_mass}
\end{center}
\end{figure}

In Fig.\ref{fig:dN_mass}, $d_N$ for neutron and proton are plotted as a function of pion mass squared, 
$m_{\rm PS}^2$, together with the quenched results. 
Unfortunately,  because of large statistical errors in full QCD results, 
it is hard to observe a difference from the quenched results.
Compared with the model calculation \cite{model_CA}, the central value is 10 times larger 
although its errors are large.

\begin{figure}[h]
\begin{center}
\vskip 10mm
\includegraphics[width=85mm, angle=0] {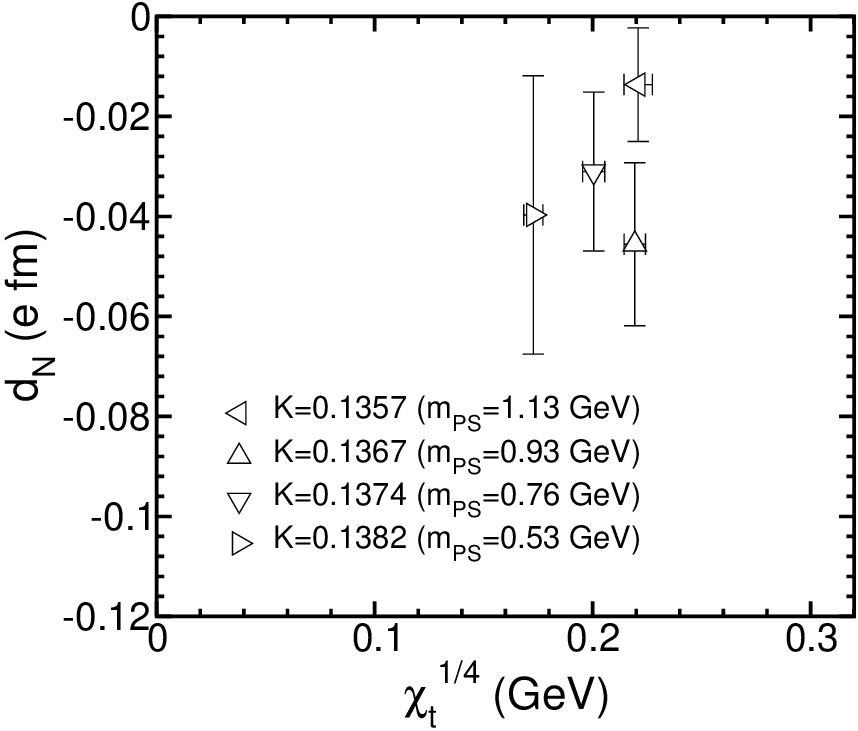}
\caption{Neutron EDM as a function of topological susceptibility $\chi_t$.} 
\label{fig:dN_Q}
\end{center}
\end{figure}

Mass dependence of $d_N$ in full QCD does not show the expected decrease toward the chiral limit. 
There are several possible explanations. 
Firstly, large statistical errors might hide the actual decrease 
of $d_N$ toward the chiral limit. 
Secondly, the quark mass in this full QCD simulation is still too heavy 
to see the decrease. 
Thirdly, $d_N$ does not vanish in the chiral limit
due to the lattice artifact that the chiral symmetry is explicitly broken in the Wilson-type quark action.
Indeed the topological susceptibility,
which is also theoretically expected to vanish in the chiral limit, 
does not show the decrease in full QCD configurations of this paper\cite{cppacs_Q}.
In Fig.\ref{fig:dN_Q}, $d_N$ is plotted as a function of the
topological susceptibility instead of the pion mass squared. 
Contrary to the case of the pion mass squared,
the change of the topological susceptibility is too little to
observe a possible decrease of $d_N$ with large statistical errors.
This suggests that
the combination of the 2nd and 3rd possibilities is a main reason for 
the mass dependence of $d_N$ in Fig. \ref{fig:dN_mass}.

\begin{figure}[h]
\begin{center}
\vskip 10mm
\includegraphics[width=85mm, angle=0] {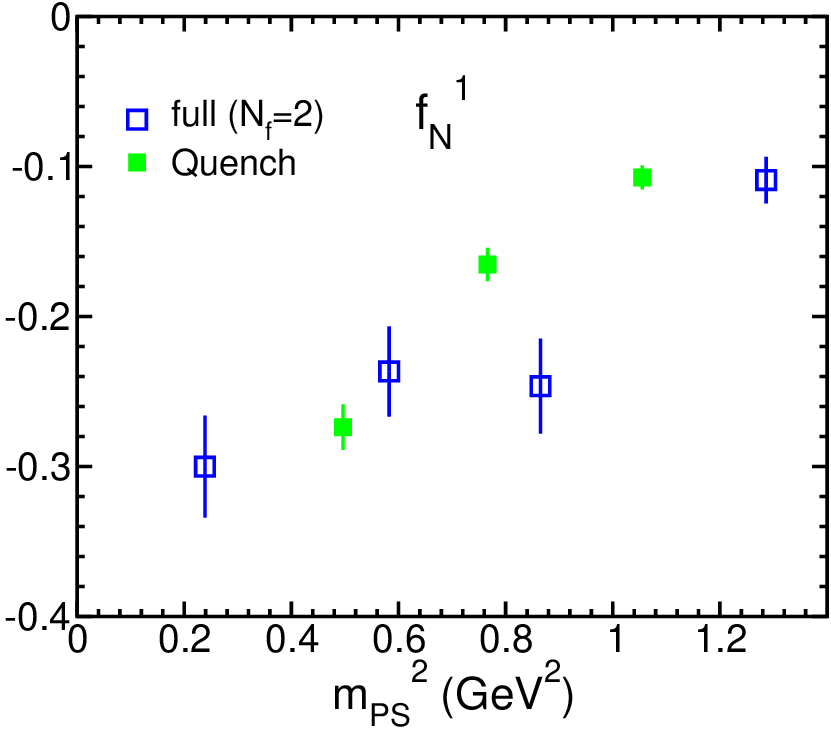}
\caption{CP-odd phase factor of the nucleon propagator $f_N^1$ as a function of the pion mass squared.} 
\label{fig:f_N}
\end{center}
\newpage
\end{figure}

We also present the results of the mass dependence 
for the nucleon CP-odd phase factor in full QCD calculation.
As discussed in \cite{Shintani_form,Shintani_wE}, the next-leading term in the nucleon spinor structure 
contains additional phase factor $f_1^N$, which arises from 
$m^\theta_Ne^{if^1_N\theta\gamma_5}=m_N(1+if_N^1\theta\gamma_5)+\mathcal O(\theta^2)$, as
\begin{equation}
  \langle N(\vec{p},t) \bar N(\vec{p},0) Q\rangle  =
  \vert Z_N\vert^2 e^{-E_{N} t}\frac{ f_N^1m_{N}}{2E_{N}} \gamma_5.
\end{equation}
where $\vec p,\,Z_N,\,E_N$ denotes the nucleon momentum, amplitude and energy, respectively. 
This factor should goes to zero toward the massless limit because of the same reason as NEDM does.
In Fig. \ref{fig:f_N}, $f_N^1$ in full QCD is plotted as a function of $m_{\rm PS}^2$, together with
the quenched results.
Although the statistical errors are not so large compared to the EDM results, 
it does not show the expected decrease toward the chiral limit as well as
the EDM case.
This observation suggests that it is unlikely
that the correct chiral behavior of the EDM is hidden 
in its large statistical errors.

\section{Summary and discussion}\label{sec:5}
In this paper we present the evaluation of the NEDM in full QCD simulation using the external 
electric field method. 
After accumulating a huge number of statistics with multiple sources and an average 
over forward and backward propagation, 
we have obtained non-zero value of the nucleon EDM for the first time in full QCD.
Statistical errors of the EDM in full QCD are still larger than in the previous quenched case.
The mass dependence of the EDM, which is similar to the quenched one,
does not show the expected behavior in full QCD that it vanishes towards the chiral limit.
Besides the large statistical errors, there may be two main reasons.
One is that sea quark masses used in this paper are still  too heavy to see the expected decrease, 
the other is that the explicit chiral symmetry breaking of the Wilson type quark action
spoils the expected chiral behavior.
Indeed this chiral behavior of the EDM is very similar to that of the topological susceptibility. 

Since it now becomes possible to calculate the EDM in full QCD, we should proceed 
toward the precise evaluation of the NEDM.
First of all, we should further decrease the sea quark mass to clearly observe the chiral behavior of EDM.
As reported in the recent advanced work in PACS-CS collaboration \cite{pacscs}, 
$2+1$ flavors dynamical configurations are being generated at 
the quark mass close to the physical point, and these configurations will be available soon.
Secondly, we should decrease statistical errors of the EDM. 
For this purpose, it may be better to switch to the form factor method \cite{Shintani_form}, 
though the zero momentum transfer limit has to be taken in this case.
From the previous work \cite{Shintani_wE} and this work, the good chiral behavior of the quark action
seems not so relevant to obtain the signal of the EDM, 
and we are currently calculating the EDM form factor using the clover quark action \cite{Shintani_form2}.
Finally the effect of disconnected loop diagram, ignored in our studies, should also be included
in the calculation. 
A preliminary study shows that it is possible to include this effect 
with the current available resources \cite{Shintani_form2}. 

\section*{Acknowledgments}
This work is supported in part by Grant-in-Aid of the Ministry of Education
(Nos.
13135204, 
15540251, 
18540250, 
70447225  
).
Numerical simulations are performed on Hitachi SR11000
at High Energy Accelerator Research Organization (KEK).
At KEK this simulation is under support of Large Scale Simulation Program (No. 06-04).
We use the two-flavor full QCD configurations generated 
by CP-PACS collaboration \cite{cppacs}.


\newpage
\begin{table}
\caption{Table for the simulation parameters in this work. 
$\mathcal O(a)$ improved coefficient of the clover term $c_{SW}$ is chosen as $c_{SW}=1.47$.
The column of $(A,B)$ denotes the smeared source parameters, 
and $K$ denotes the hopping parameter for the degenerate up and down quarks.}
\label{tab:lat_param}
\begin{center}
\begin{tabular}{ccccccccc}
\hline
\hline
  Lattice size & Physical volume (fm${}^3$) & cutoff $a^{-1}$ (GeV) & $K$ 
  & $(A,B)$ & $m_{PS}/m_V$ & $m_Na$ \\
\hline
  $24^3\times 48$ & 2.6$^3$ & $1.83$ & 0.1357 & (1.5,0.45) & 0.81 & 1.1851(10)\\
                  &         &        & 0.1367 & (1.5,0.43) & 0.76 & 1.0224(11)\\
                  &         &        & 0.1374 & (1.5,0.35) & 0.69 & 0.8944(12)\\
                  &         &        & 0.1382 & (1.5,0.25) & 0.58 & 0.7167(14)\\
\hline
\hline
\end{tabular}
\end{center}
\end{table}
\begin{table}
\caption{Results for neutron and proton EDM at each quark mass}
\label{tab:EDM}
\begin{center}
\begin{tabular}{ccccccccc}
\hline
\hline
 $K$ & $m_{PS}$ (GeV) & neutron EDM (e$\cdot$fm) & proton EDM (e$\cdot$fm)\\
\hline
0.1357 & 1.13 & -0.014(11) & 0.049(21) \\
0.1367 & 0.93 & -0.046(16) & 0.019(29) \\
0.1374 & 0.76 & -0.031(16) & 0.060(28) \\
0.1382 & 0.53 & -0.040(28) & 0.072(49) \\
\hline
\hline
\end{tabular}
\end{center}
\end{table}
\end{document}